\documentclass[aip,jap]{revtex4-1}

\usepackage{graphicx,amsmath,amssymb,subfig}

\preprint{AIP/123-QED}

\begin{document}
\title{Spin relaxation due to electron-electron magnetic interaction in high Lande $g$-factor semiconductors} 
\author{Akashdeep Kamra}\email{akashk@iitk.ac.in}
\author{Bahniman Ghosh} 
\affiliation{ Department of Electrical Engineering, IIT - Kanpur, Kanpur - 208016, India}
\author{Tarun K. Ghosh}   
\affiliation{Department of Physics, IIT - Kanpur, Kanpur - 208016, India}

\date{\today}

\begin{abstract}
We investigate spin transport in InSb/InAlSb heterostructure using the Monte Carlo approach, generalized by including density matrix description of spin for taking spin dynamics into account. In addition to the dominant Dyakonov-Perel (DP) mechanism for spin control and relaxation, we consider magnetic interaction between electrons which assumes importance due to high electronic Lande $g$-factor in the material. It is found that while the effect of magnetic interaction is not important at low densities, it reduces the spin relaxation length by as much as $50 \%$ at higher densities. We also present a calculation which elucidates the suppression of decoherence attributed to wave vector dependent magnetic field in the DP relaxation mechanism. We note that magnetic interaction is a general relaxation mechanism which may assume importance in materials with high electronic Lande $g$-factor.
\end{abstract}


\maketitle

\section{Introduction}
The electron spin dynamics has been studied extensively due to the prospects spin based devices hold for future. In addition to offering improvements over the contemporary devices, Spintronics promises devices with entirely new functionalities \cite{fabian,bandyopadhyay,zutic}. While metal based Spintronics has grown into a somewhat mature field over the past few years owing to several succesful applications of magnetic tunnel junctions, semiconductors still remain to make an impression. The importance of integrating spintronics with contemporary semiconductor technology cannot be overstated.

Various \textrm{III-V} compounds are being investigated for the material to be used in semiconductor spintronics. The narrow gap semiconductors InSb and InAs offer distinct advantages over their wide gap counterparts \cite{gilbertson}. Spin transport in both these semiconductors has been studied to find InSb better of the two \cite{litvinenko}. InSb has the lowest effective mass amongst \textrm{III-V} semiconductors, high mobility, and a stronger spin-orbit coupling. Further, the Rashba \cite{rashba} term has been found to dominate the Dresselhaus \cite{dresselhaus} term in the spin-orbit interaction at high densities \cite{gilbertson,branford}. This, in principle, must offer a better control over the spin dynamics in InSb as Rashba coupling can be tuned by electric field. These factors make InSb a very promising candidate as semiconductor material for Spintronic applications in general, and for Datta-Das transistor \cite{datta} in particular.

Of the three spin relaxation mechanisms in semiconductor systems, Bir-Aronov-Pikus mechanism \cite{bap} is relevant for p type semiconductors only and Elliot-Yafet mechanism \cite{ey} is dominated by Dyakonov-Perel (DP) mechanism \cite{dp} in high mobility samples \cite{litvinenko,privman}. We propose yet another pathway to spin relaxation that gains importance in semiconductors with high $g$-factor. The electronic magnetic moment gives rise to an actual magnetic field which plays an important role in addition to the effective magnetic field attributed to spin-orbit coupling. This effect is proportional to $g^{2}$ and is three orders of magnitude higher in InSb ($g \sim -50$) as compared to other materials of interest such as GaAs ($g \sim -0.44$) \cite{gfact}. The motivation for considering magnetic interaction is derived from the observation that in ferromagnets, magnetic interaction leads to a phenomenon as important as the formation of domain walls. Hence, one is inclined to believe that magnetic interaction ought to play some role in restoring the spin space equilibrium in non magnetic systems for which $g^{2}$ becomes comparable to relative permeability in typical ferromagnets.

In this paper, we study the effect of electron-electron magnetic interaction on spin transport in high $g$-factor semiconductors using Monte Carlo simulations \cite{privman,jacoboni,lugli}. Owing to the reasons discussed above, we study InSb/InAlSb heterostructure as a model system. However, any given system can be simulated using our routine. The spin dynamics in the system, owing to its high mobility, is dominated by DP mechanism. We calculate the magnetic field attributed to non equilibrium spin concentration in a mean field approximation by assuming uniform distribution of charge and spin along the width of the channel. This magnetic field adds up to the pseudo field attributed to spin-orbit coupling to produce the net field experienced by electrons. We have obtained dependence of spin relaxation length on electron density. A similar behavior has been observed in experimental studies on InSb \cite{kallaher}. In the section on results, we present a calculation which throws some light on the need to include magnetic interaction effect in addition to DP mechanism. Finally, we conclude with a discussion on the requirement of a more accurate calculation than the mean field approach considered in this work.

\section{Model}
A detailed account of the Monte Carlo method used for studying spin dynamics in \textrm{III-V} semiconductors has been given elsewhere \cite{privman,jacoboni,lugli}. We shall only enlist the essential features of the simulation. The dynamics has been assumed to be confined to the lowest subband. We treat the band structure via the nonparabolic approximation due to Conwell and Vassel \cite{vassel,jacoboni}. 
\begin{equation}
E(1 + \alpha E) = \frac{\hbar^{2}k^{2}}{2m}
\end{equation}
Here, $\alpha$ is given by the following expression \cite{fawcett}.
\begin{equation}
\alpha = \frac{1}{E_{g}} \left(1 - \frac{m}{m_{0}}\right)^{2}
\end{equation}
$E_{g}$ is the bandgap, $m$ the effective mass of electron and $m_{0}$ is the mass of free electron. The scattering mechanisms incorporated in the simulation are acoustic phonon scattering, optical phonon scattering and ionized impurity scattering. The ionized impurity scatteing calculation considers the Conwell and Weisskopf approach \cite{conwell,lugli}. We now discuss the boundary condition used in the system. In a realistic situation, there is a constant flux of electrons, independent of the applied voltage, from both electrodes. At zero voltage and steady state, the inward flux from electrodes is exactly balanced by the outward flux from the channel. We exploit the fact that the outward flux from the channel changes marginally on application of a reasonable voltage. Hence, we choose to reintroduce the electrons, that fly into source, at the drain end and vice versa. This boundary condition distorts the spin polarization profile from exponential decay for a small channel device and presents a realistic picture, which  allows us to take into account effects like spin selective transmission of electrons across drain boundary.

\begin{figure}[hbt]
\subfloat[][]
{\includegraphics{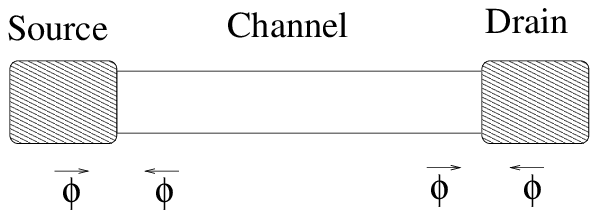}}
\qquad
\subfloat[][]
{\includegraphics{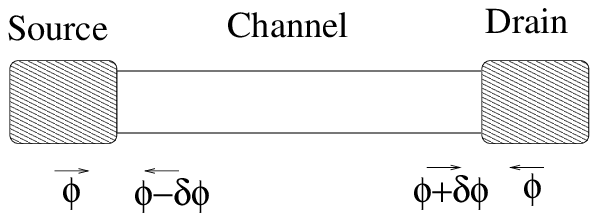}}
\\
\subfloat[][]
{\includegraphics{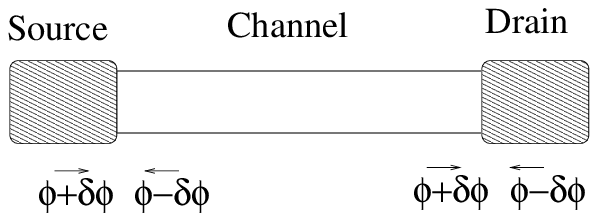}}
\caption{Depiction of electronic flux in steady state at the two electrodes, Source and Drain, at (a) zero voltage and (b) small positive voltage $V_{DS}$. Case (c) depicts the approximation that we consider in our calculations. Since $ \delta \phi $  is very small, cases (b) and (c) are nearly the same.}
\end{figure}

It is not possible to compute the complete 2D magnetic field intensity variation across the device for a given distribution of electrons at each simulation step. Hence, we adopt a mean field kind of approach towards its calculation and ignore any variation in magnetic field along the width of the device. For the purpose of simulation, the device is divided into a grid along the length of the channel. The magnetic field is calculated in two parts - interaction within the grid element and interaction between different grid elements. Interaction within the grid element is calculated assuming all the electrons in it to be spaced uniformly along width of the device and the net magnetic moment to be distributed equally amongst them.
\begin{equation}\label{intragrid}
 B^{intragrid}_{x,y,z} \ = \ (-1_{x},2_{y},-1_{z}) \ \frac{\mu_{0}}{4\pi} \ \frac{M_{x,y,z}N^{2}}{W^3} \ 2 \sum_{1}^{\infty} \frac{1}{n^3} 
\end{equation}
Coordinate system has been chosen such that $\hat{x}$ and $\hat{y} $ are along length and width of the device respectively.  Here $(-1,2,-1)$ corresponds to multiplicative factors for $x$,$y$ and $z$ components respectively. $N$ denotes the total number of electrons in the grid element, $M_{x,y,z}$ denotes respective components of total magnetic moment due to all the electrons in the grid element and $W$ is the width of the device.

For calculation of inter grid element interaction, magnetic field due to each grid element, at other grid elements, is calculated assuming the net magnetic moment to be distributed uniformly along width of the device. It is a simple task of evaluating some integrals to find the required magnetic field intensity. Ignoring edge effects, magnetic field intensity due to a grid element 1 at a grid element 2 is given by the following expression.
\begin{equation}
B^{intergrid}_{x,z} \ = \ (1_{x},-1_{z}) \ \frac{\mu_{0}}{2\pi} \ \frac{M_{x,z}}{W} \ \frac{1}{r_{12}^{2}}
\end{equation}
Symbols have their usual meanings (See Eq. (\ref{intragrid})) and $r_{12}$ is the distance between the two grid elements. It is observed that $y$ component of intergrid interaction vanishes. The expressions above enable us to calculate magnetic field intensity along the channel length for a given spin polarization profile, which combined with spin-orbit coupling dictates spin dynamics of the electrons.

In the range of densities considered in this work, Rashba term dominates spin-orbit coupling \cite{gilbertson}. Hence, a major contribution to spin dynamics comes from the following Rashba Hamiltonian.
\begin{equation}
H_{R} \ = \ \alpha \left(k_{y}\sigma_{x} - k_{x}\sigma_{y} \right)  
\end{equation}
We consider Dresselhaus term in its linear approximation, the Hamiltonian for which is as given below.
\begin{equation}\label{dreshamil}
H_{D} \ = \ \beta <k_{z}^{2}> \left( k_{y} \sigma_{y} - k_{x} \sigma_{x} \right )
\end{equation}
Addition of the Zeemann term takes care of the magnetic field intensity in a semi classical way. This method is approximate as a complete solution requires consideration of vector potential in the coordinate space Hamiltonian, which leads to further complications as free particle wave is no longer the solution. However the condition that magnetic field intensity (B) is small enough so that the momentum relaxation time is much smaller than $m/eB$ ensures that the effect of magnetic field on the coordinate space wavefunction remains minute. We deal with the effect of magnetic field in coordinate space by adding a Lorentz force term to coordinate space dynamics. 

\section{Results and Discussion}
The structure probed in simulations is similar to the one experimentally studied by Gilbertson and co workers \cite{gilbertson}. The thickness of InSb layer in the considered device structure is taken to be $20nm$ while both length and width are taken as $300nm$. The choice of longitudinal dimensions is influenced by the inability to simulate more than $2 \times 10^{5}$ electrons. The values of $\alpha$ and $\beta <k_{z}^2>$ were taken as $1.5 \times 10^{-11} eV m$ and $3 \times 10^{-12} eV m$ respectively \cite{gilbertson}. The dynamics is assumed as being confined to the lowest subband, the energy for which was calculated as $0.1 eV$. The various material and scattering parameters were adopted from a standard manual on Monte Carlo simulations \cite{damocles}. A remote doping scheme such as an adjacent doped layer allows for higher mobility and hence an effective impurity density of $2 \times 10^{11} cm^{-2}$ is assumed to obtain a reasonable value of mobility. A basic time step of $0.1 fs$ was decided and electrons were evolved for $2 \times 10^{5}$ steps to achieve a steady state. Data for time averaging was recorded for the last $10000$ steps only. The validity of simulation was thoroughly tested by reproducing already reported results \cite{privman}.

\begin{figure}[hbt]
\subfloat[][]
{\includegraphics[height=6.3cm,width=8.0cm]{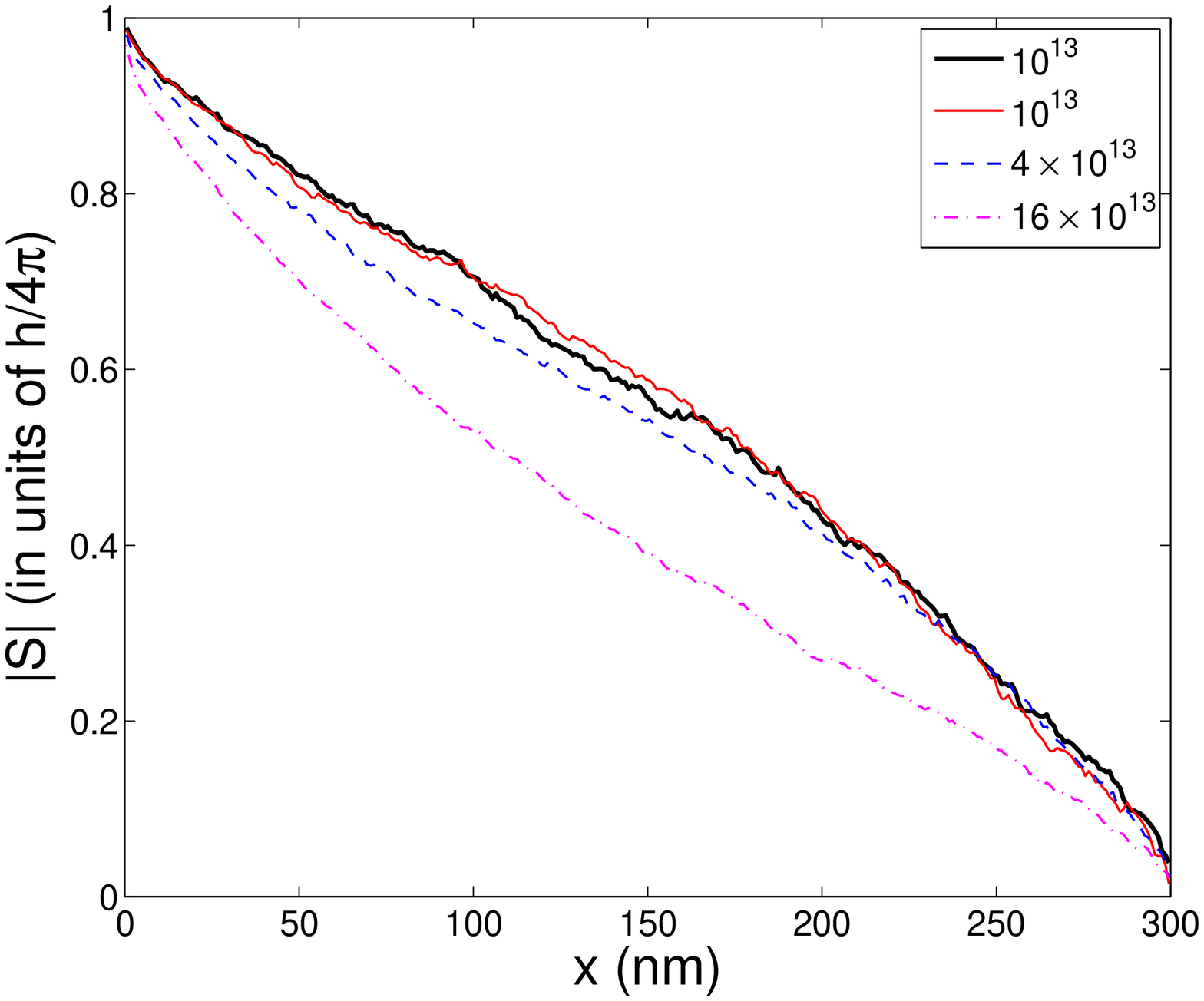}\label{Scomparison1}}
\quad
\subfloat[][]
{\includegraphics[height=6.3cm,width=8.0cm]{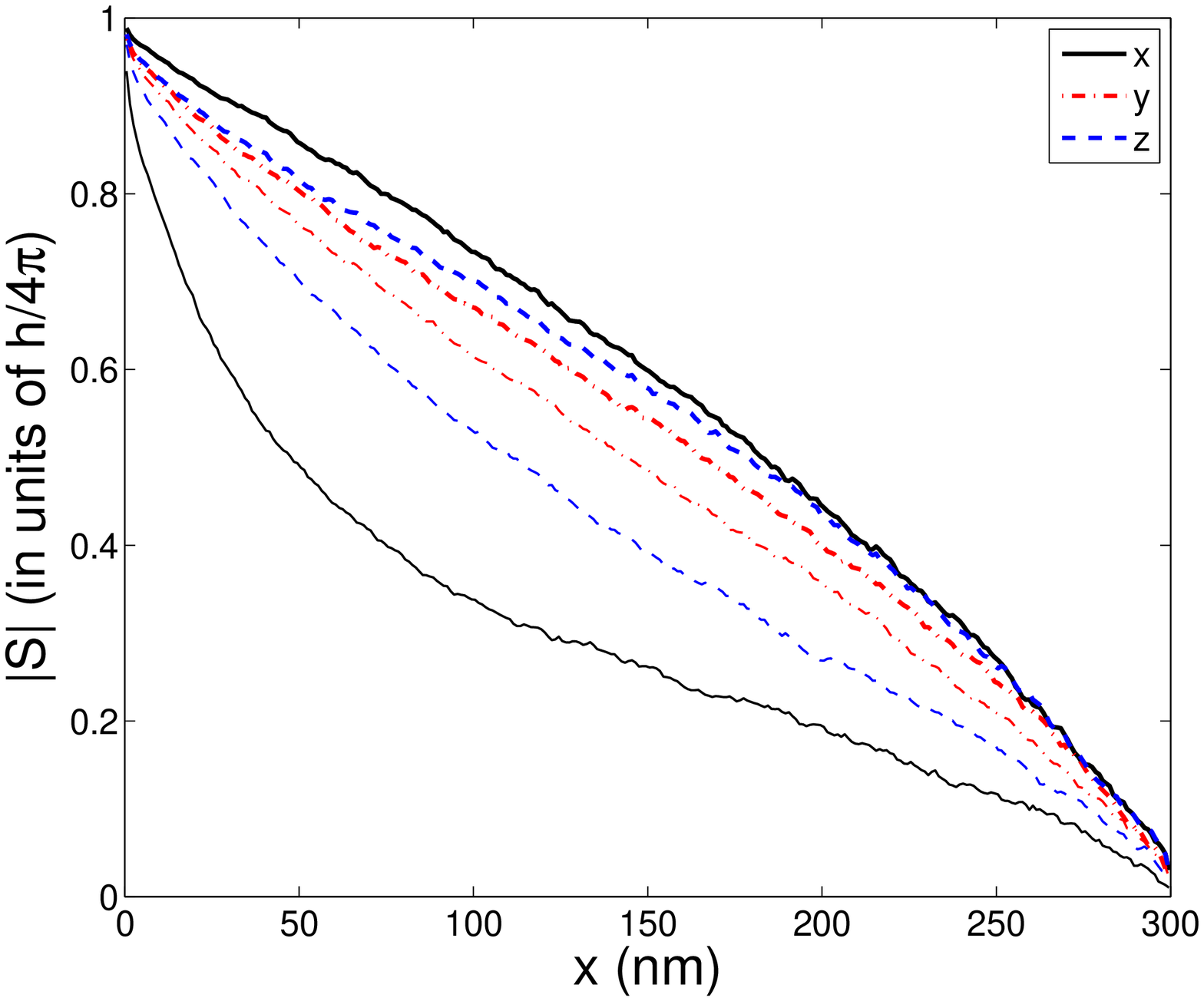}\label{Scomparison2}}
\caption{ $|S|$ along channel length. Thick lines denote the case when magnetic interaction is not accounted for in calculations. (a) Legend indicates electron density in $cm^{-2}$ and electrons are polarized along $z$ direction at the time of injection. (b) Legend indicates the polarization direction of electrons at the time of injection and $n = 16 \times 10^{13} cm^{-2}$. Maximum reduction in spin relaxation length of more than $50 \%$ is observed for the case of initial polarization along $\hat{x}$. }
\end{figure}

Spin transport was simulated assuming $100 \%$ injection efficiency with $V_{DS} = 0.05 V$ for different electron densities. Room temperature was assumed in all simulations. The results are reported in Fig. (\ref{Scomparison1}). We note that spin relaxation profile is independent of electron density when DP mechanism alone is considered. However, magnetic interaction is strongly dependent on density with its effect becoming reasonable at a density of $4 \times 10^{13} cm^{-2}$. The decrease in spin relaxation length due to magnetic interaction is almost $30 \%$ for a density of $16 \times 10^{13} cm^{-2}$. It is to be noted that data close to drain was ignored while fitting in order to nullify the boundary effect.

\begin{figure}[hbt]
\subfloat[][]
{\includegraphics[height=6.3cm,width=8.0cm]{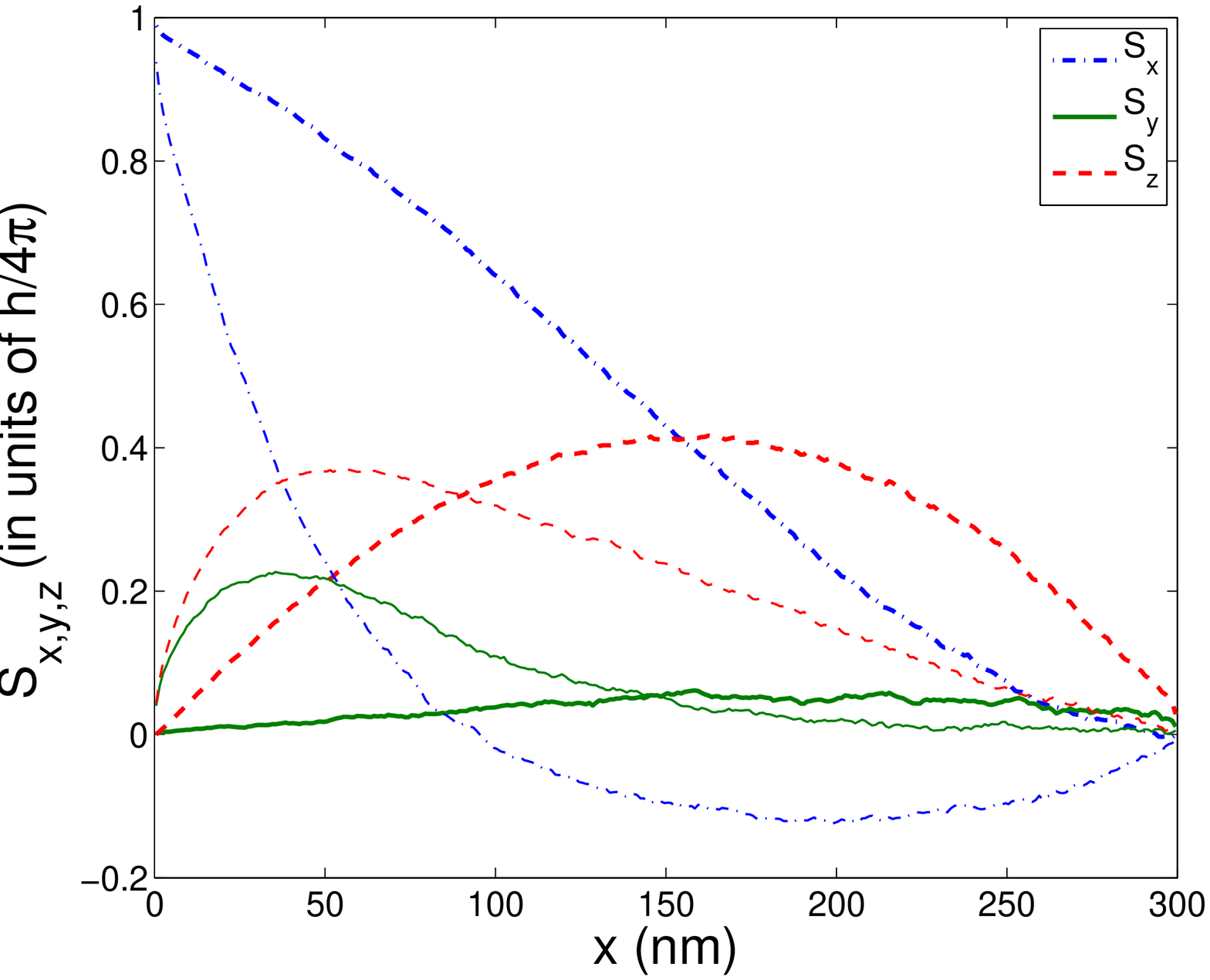}\label{evol}}
\quad
\subfloat[][]
{\includegraphics[height=6.3cm,width=8.0cm]{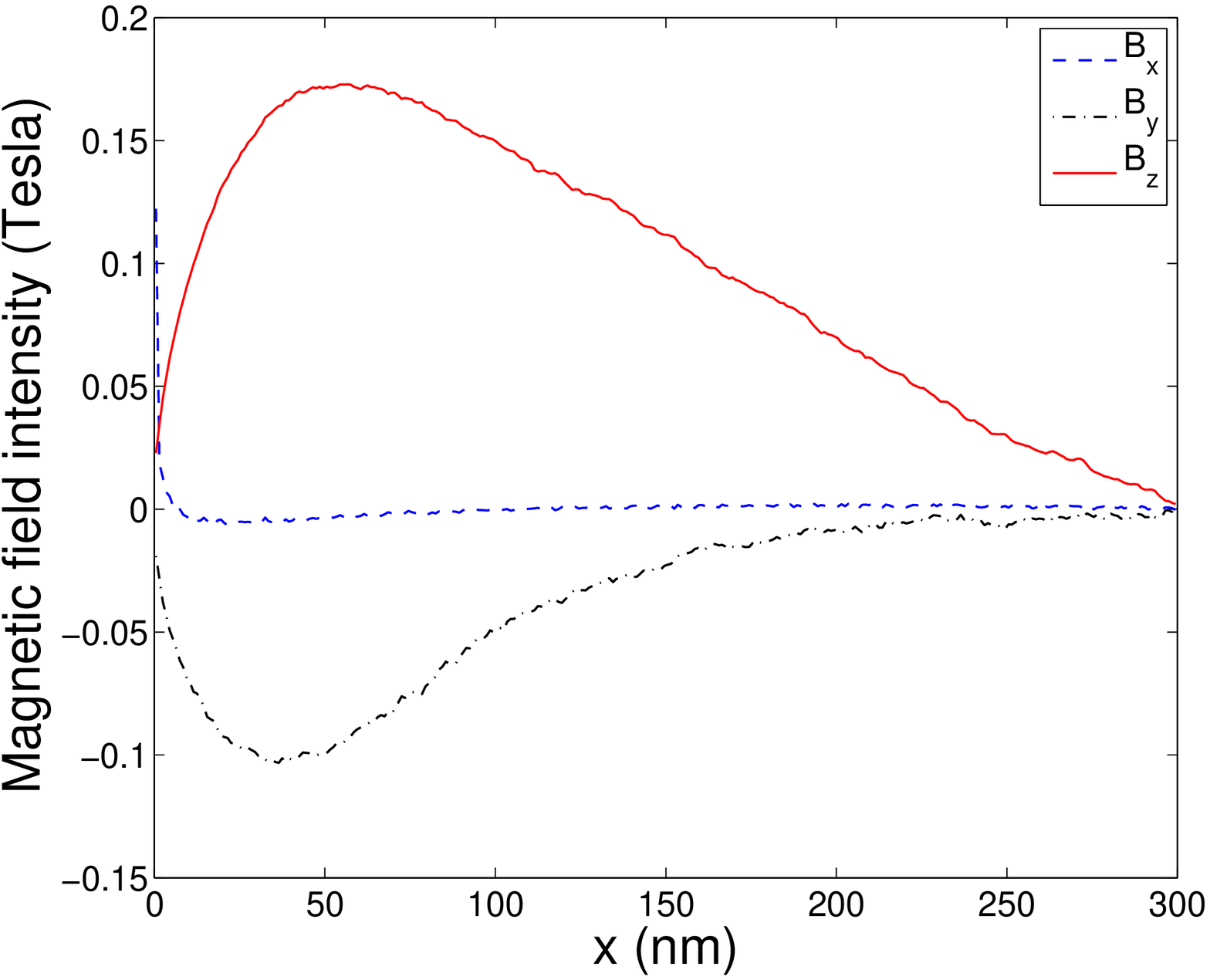}\label{Bfield}}
\caption{ Initial polarization of electrons is along $\hat{x}$ and $n = 16 \times 10^{13} cm^{-2}$. (a) $S_{x,y,z} $ along the channel length. Thick lines denote the case when magnetic interaction is not accounted for in calculations. (b) Magnetic field intensity variation along the channel length. }
\end{figure}

Anisotropic nature of the spin-orbit Hamiltonian and magnetic interaction leads to different spin dynamics for different initial polarization of injected electrons. We compare the spin dynamics, with and without taking magnetic interaction into account, in Fig. (\ref{Scomparison2}). Our simulations show that the effect of magnetic interaction is most pronounced for the case when electrons are initially polarized in $x$ direction. We compare the complete spin dynamics for this case in Fig. (\ref{evol}). We note that spin profile along the channel is modified substantially by the inclusion of magnetic interaction in our calculations.

We now discuss a calculation which brings out a somewhat diminished role of $\vec{k}$ dependent effective magnetic field in causing spin decoherence. The calculation also points to a reduced decoherence attributed to variability of electron trajectories. We show that spin-orbit coupling is more important as a control mechanism than as a decoherence mechanism, and hence the need to consider other relaxation mechanisms for accurate spin dynamics. This result is in our favor as control via spin-orbit coupling is at the heart of many spintronic devices \cite{fabian,datta}. In what follows, we ignore the weak Dresselhaus term (Eq. (\ref{dreshamil})) at first and comment on its significance later. Considering the spin-orbit coupling alone, the spin dynamics is given by the following equations.
\begin{eqnarray}
\frac{d \sigma_{x}}{dt} \ & \ = \ & \ -\frac{2\alpha}{\hbar} k_{x} \sigma_{z} \label{sigmaxevol} \\
\frac{d \sigma_{y}}{dt} \ & \ = \ & \ -\frac{2\alpha}{\hbar} k_{y} \sigma_{z} \label{sigmayevol}  \\
\frac{d \sigma_{z}}{dt} \ & \ = \ & \ \frac{2\alpha}{\hbar} \left( k_{y} \sigma_{y}  + k_{x}  \sigma_{x} \right) \label{sigmazevol}
\end{eqnarray}
In the equations above, it is deemed as understood that $\sigma_{x,y,z}$ (which stand for Pauli matrices) and $k_{x,y,z}$ represent their respective expectation values. Let us focus our attention on Eq. (\ref{sigmaxevol}). Writing $k_{x}$ in terms of $ v_{x}$, integrating by parts and substituting $d \sigma_{z}/dt$ from Eq. (\ref{sigmazevol}),
\begin{equation}
\sigma_{x} \ = \ \sigma_{x}^{0} - \frac{2\alpha m}{\hbar^2} \left[ \sigma_{z} x \ - \ \frac{2\alpha m}{\hbar^2} \int_{0}^{t} \left( v_{y} \sigma_{y} + v_{x} \sigma_{x} \right) x dt \right]
\end{equation}
In writing the above expression, it was assumed that at $t = 0$ electron was introduced at $ x = 0 $ with its initial polarization in the $x$-$z$ plane. $\sigma_{x}^{0}$ and $\sigma_{z}^{0}$ represent respective initial values. Now, integrating Eq. (\ref{sigmazevol}) and substituting for $\sigma_{z}$ in the above equation,
\begin{equation}\label{sigmaxfinal}
\sigma_{x} \ = \ \sigma_{x}^{0} - \frac{2\alpha m x \sigma_{z}^{0}}{\hbar^2} \ - \ \frac{4\alpha^{2} m^{2}}{\hbar^{4}} \left[ x \int_{0}^{t} \left( v_{y} \sigma_{y} + v_{x} \sigma_{x} \right) dt - \int_{0}^{t} x \left( v_{y} \sigma_{y} + v_{x} \sigma_{x} \right) dt \right]
\end{equation}
We pause here to build the background for discussion on the significance of expression derived above. Firstly, we note from simulation results (Fig. (\ref{evol})) that there is little change in $\sigma_{y}$ from its initial value. This is a consequence of a strong electric field along the $x$ direction which results in $|k_{x}| > |k_{y}|$ on an average, which causes the effective magnetic field due to spin-orbit interaction to be almost aligned along $y$ direction. Hence, we are justified in treating $\sigma_{y}$ as almost constant. Since $\sigma_{x}^{2} + \sigma_{y}^{2} + \sigma_{z}^{2} = 1$ and with $\sigma_{y}$ fixed, only one of $\sigma_{x}$ and $\sigma_{z}$ is independent. Now we are ready to discuss Eq. (\ref{sigmaxfinal}). We note that the first two terms on the right hand side are same for all electrons irrespective of their velocities or the paths they take to reach a particular $x$. Hence, the $\vec{k}$ dependent Hamiltonian does not lead to decoherence to the first order in spin-orbit coupling parameter $\alpha$. The first order term provides us with the much needed control of spin dynamics through spin-orbit interaction. The second order term, which is relatively small, is the one responsible for spin decoherence.

\begin{figure}[hbt]
\includegraphics{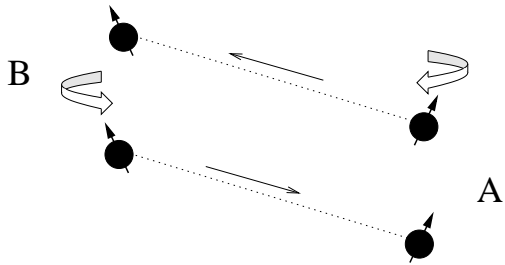}
\caption{Physical demonstration of coherent first order term.} \label{demo}
\end{figure}

Physically, this can be understood with the help of a simplified example. Consider an electron that moves from a point A to another point B and back to point A (Fig. (\ref{demo})). As the electron moves from A to B, it undergoes precession about an effective magnetic field and thus suffers a change in its polarization. Now as it reverses the direction of its motion to move back to point A, the direction of effective magnetic field is reversed as well and the electron begins to precess in the opposite direction. The faster it moves, the faster it precesses so that when it comes back to point A, it has undergone almost the same change in polarization in the reverse direction. On the other hand, magnetic interaction does not have any coherent effect as the precession angle depends upon the path traversed by the electron, which is different for all of them. Further, the magnetic field is stronger in the region with higher spin polarization.

It should be noted that inclusion of Dresselhaus term in this calculation yields an additional term $\beta <k_{z}^{2}> y$ added to $\alpha x$ in the first order coherent term of Eq. (\ref{sigmaxfinal}). The dependence of polarization on $y$ coordinate can lead to substantial decoherence after summing up over all the electrons at a given $x$. This result elucidates the advantage in suppressing the importance of Dresselhaus term for better control of spin dynamics in the device. Hence, using a high electron density in the device might be useful for better spin control \cite{gilbertson} in addition to increasing the speed.

We conclude this section by emphasizing the need for a better 2D model than the mean field model considered for magnetic interaction in this work. In our calculations, all the electrons in a grid element experience the same magnetic field irrespective of their $y$ coordinate and their distribution within the grid element. This is certainly not true and leads to an underestimation of the actual decoherence rate. As a result, in our simulations, the density at which magnetic interaction effects gain importance comes out to be somewhat high. In actual devices, magnetic effects in InSb may set in at much lower electron densities. The reported magnetic effects can be responsible for the observed dependence of spin decoherence rate on electron density in InSb \cite{kallaher}.

\section{Conclusion}
We have discussed a new pathway to spin decoherence in high $g$-factor semiconductors. Our simulations show an increased spin decoherence attributed to electron-electron magnetic interaction. We have carried out an approximate calculation which elucidates the weakness of DP spin relaxation mechanism in the context of Datta-Das transistor. It also manifests the advantage of a diminished importance of Dresselhaus term in the overall spin-orbit interaction. InSb appears to be an appealing candidate for semiconductor based spintronics due to low electron effective mass and relatively strong spin-orbit coupling. At the same time, a high Lande $g$-factor brings about additional effects. The electron-electron magnetic interaction leads to a shorter spin relaxation length. Hence, InSb might not be a suitable material for spin transport devices.

\section*{Acknowledgement}
We thank Mr. Sahil Suneja for his help with simulations. We thank Prof. M. K. Harbola of IIT Kanpur for useful discussions. This work was funded by Ministry of Human Resource Development (MHRD), India.

\end{document}